\documentclass[conference]{IEEEtran}
\IEEEoverridecommandlockouts
\usepackage[T1]{fontenc}
\usepackage{threeparttable}  
\usepackage[misc,geometry]{ifsym}
\usepackage[linesnumbered,ruled]{algorithm2e}
\usepackage{amsmath}
\usepackage{threeparttable}
\usepackage{subcaption}
\usepackage{multirow}
\usepackage{graphicx}

\usepackage{amsfonts}
\usepackage{graphicx}
\def\BibTeX{{\rm B\kern-.05em{\sc i\kern-.025em b}\kern-.08em
    T\kern-.1667em\lower.7ex\hbox{E}\kern-.125emX}}
\begin{document}

\title{Medical Image Denosing via Explainable AI Feature Preserving Loss
}

\author{\IEEEauthorblockN{Guanfang Dong}
\IEEEauthorblockA{\textit{Multimedia Research Center} \\
\textit{University of Alberta}\\
guanfang@ualberta.ca}
\and
\IEEEauthorblockN{Anup Basu}
\IEEEauthorblockA{\textit{Multimedia Research Center} \\
\textit{University of Alberta}\\
basu@ualberta.ca}
}

\maketitle

\begin{abstract}
	Denoising algorithms play a crucial role in medical image processing and analysis. 
	However, classical denoising algorithms often ignore explanatory and critical medical features preservation, which may lead to misdiagnosis and legal liabilities.
	In this work, we propose a new denoising method for medical images that not only efficiently removes various types of noise, but also preserves key medical features throughout the process.
	To achieve this goal, we utilize a gradient-based eXplainable Artificial Intelligence (XAI) approach to design a feature preserving loss function.
	Our feature preserving loss function is motivated by the characteristic that gradient-based XAI is sensitive to noise.
	Through backpropagation, medical image features before and after denoising can be kept consistent.
	We conducted extensive experiments on three available medical image datasets, including synthesized 13 different types of noise and artifacts.
	The experimental results demonstrate the superiority of our method in terms of denoising performance, model explainability, and generalization.
	
\end{abstract}

\section{Introduction}\label{intro}
Medical image denoising is an important stage in medical image processing and analysis.
Its main objective is to remove noise from a medical image while retaining key information about lesions and structures \cite{kaur2023complete}.
Unlike conventional image denoising algorithms, medical image denoising requires a judicious and transparent approach to noise removal.
The improper or unexplained removal of noise may lead to misleading diagnosis, lack of trust, and legal and ethical liabilities \cite{van2022explainable}.
Therefore, we are motivated to propose a new method that effectively removes all types of noise, while preserving key medical image features and providing explainability.

In order to preserve critical medical features during the denoising process, we leverage the potential of eXplainable Artificial Intelligence (XAI).
XAI is a broader research topic that aims to improve the transparency, explainability, and trustworthiness of AI systems.
In this work, we adopt a gradient-based XAI approach to extract the medical image features. 
In general, gradient-based XAI explains the model behaviour by exploiting gradient information during backpropagation of deep neural networks.

In this work, we employ a deep learning model for image restoration task, represented by a function $f$. 
This function maps an input image $x$ to a restored output image $y$: $y = f(x)$. 
Gradient-based XAI is used identify which regions of the input image $x$ contribute the most significantly to the output image $y$.
To achieve this, we can compute the gradient of the model output $y$ with respect to the input $x$, denoted as $\nabla x$. 

Since the model output is a continuous value, the elements of the gradient $\nabla x$ reflect the contribution of each component in the input image $x$ to the restored output image $y$. 
A larger gradient value indicates a greater influence of the input component on $y$. 

Gradient-based XAI methods have some limitations; one of the drawback is their sensitivity to noise. 
The explanation result may be affected by noise due to the sensitivity of gradients to small changes in the input, leading to misleading explanations.
In many cases, this property is perceived as a drawback that gradient-based XAI methods must overcome. 
Nevertheless, we find that this attribute provides a potential for feature preservation in denoising networks.
Due to this limitation, the features of a clean medical image and a noisy medical image differ from each other.
Consequently, we can design a loss function that guides the network to fit the features of a clean medical image.
It is important to highlight that the gradient features obtained for clean and noisy images can be visualized using a saliency map.
This introduces explainability to our approach. 
Although the features extracted by XAI slightly differ from the actual medical image features, we are still able to obtain explanations for the model's denoising behavior.
To some extent, the trust problem associated with black-box neural network models is alleviated.

Our method is robust to different types of noise.
Furthermore, our method demonstrates the capability to generalize across various medical image acquisition devices.
To validate these claims, we conducted extensive experiments. 
Initially, our denoising experiments employed three publicly available datasets. 
They are the Lung Image Database Consortium image collection (LIDC-IDRI), RSNA Pneumonia Detection, and LiTS (Liver Tumor Segmentation), covering both CT and X-ray image acquisition devices.
At the same time, we simulated 13 different types of noise or artifacts, including Gaussian noise, Poisson noise, speckle noise, non-central chi-square noise, Rayleigh noise, salt-and-pepper noise, structure noise, thermal noise, magnetic field inhomogeneity noise, chemical shift artifact, motion artifact, wrap artifact, and susceptibility artifact.
Both the denoising and artifact removal results demonstrated that our method achieves good denoising performance.

Overall, our contributions are as follows:

\begin{enumerate}
	\item We propose a feature preserving loss that utilizes the potential of gradient-based XAI method. We leverage its sensitivity to noise to design a loss function. This loss function can guide the denoising network to fit the features of clean medical images.
	\item The feature preserving loss improves the explainability of the denoising network, allowing users to visualize the medical features before and after the denosing is applied.
	\item We perform extensive denoising experiments on large-scale medical imaging datasets and 13 different noise type. The experimental results demonstrate the robustness and generalizability of the feature preservation loss. The importance of preserving important information is thus proved.
\end{enumerate}

\section{Related Work}

Denoising is a fundamental processing step in image processing. 
The primary objective of denoising algorithms is to remove image noise while preserving critical features within the image, such as edges and textures.
Many well-known denoising algorithms aim to address this issue.
Mean Filtering \cite{castleman1996digital} removes noise by substituting each pixel's value with the average value of its neighboring pixels.
Median Filtering \cite{huang1979fast} alleviates noise by replacing the current pixel value with the median value among its neighbors.
Gaussian Filtering \cite{canny1986computational} operates in the spatial domain, utilizing Gaussian functions as weights in convolution.
Wiener Filtering \cite{kailath1981lectures} is an optimal filtering technique that minimizes the square of the error.
Wavelet Transform employs multi-scale approaches for denoising \cite{donoho1995adapting}.

However, in medical image denoising, classical algorithms encounter multiple challenges.
To efficiently eliminate noise, classical algorithms may remove both subtle structures and important features within the image, which is unacceptable in medical diagnostics.
Also, denoising medical images necessitates higher computational complexity.
Therefore, algorithms with high computational complexity, such as BM3D \cite{dabov2007image}, struggle to applied on high-resolution medical images.
Then, different types of medical images (e.g., MRI, CT, X-ray, etc.) have different noise distributions. 
This requires denoising algorithms to tune the parameters manually.
Unfortunately, most traditional algorithms lack adaptivity.
Since noise has a complex dependence on the signal, this makes linear denoising methods ineffective.

Therefore, in recent years, more research has focused on medical image denoising using machine learning and deep learning methods. 
Although learning-based methods have obvious advantages in adapting to different noise models, they also introduce new problems such as overfitting and poor interpretability.
Zhang et al. \cite{zhang2021plug} proposed a CNN denoiser prior, called DPIR, that can handle various noise levels. DPIR is applicable to interpolative image restoration methods.
Neshatavar et al. \cite{neshatavar2022cvf} proposed a self-supervised denoising method called CVF-SID. 
CVF-SID is based on a cyclic multivariate function module and a self-supervised image disentanglement framework.
However, although these recent works have considered different noise distributions, they do not consider feature extraction for noisy images.
Especially in medical image denoising, the lack of feature consideration will result in over-smoothing of the denoised images.
For this reason, Dong et al. \cite{9353537} proposed FDCNN, which uses medical features as guidance and combines them with the denoised image through fusion.
However, although their method successfully preserves medical features, it cannot eliminate noise in the feature areas.
Their work motivates us to propose a method that not only preserves the features but also denoises the feature areas.

\section{Methodology}

\begin{figure*}[tb]
	\includegraphics[width=\linewidth]{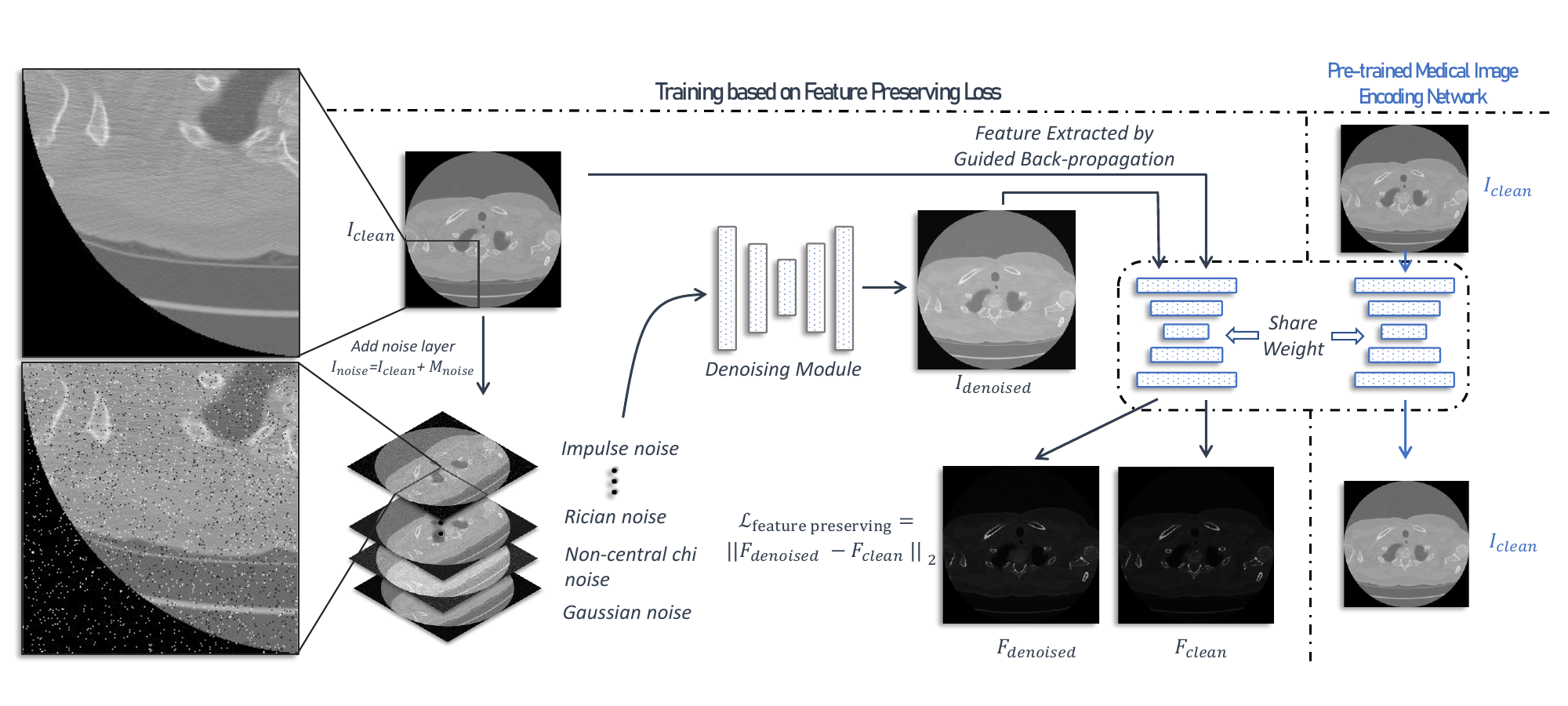}
	\caption{\label{pipeline}
		Workflow of the Proposed Method.}
\end{figure*}

As shown in Fig. \ref{pipeline}, we propose a multi-stage medical image denoising method that combines image restoration, explanation and denoising.
Our method first uses a pre-trained image restoration network. 
This network maps the noise-free medical images $I_{\text{clean}}$ to the same noise-free image space.
Next, a gradient-based explainable artificial intelligence (XAI) technique, Guided Back-Propagation \cite{springenberg2015striving}, is used to explain the output of the image restoration network.
After the pre-training image restoration network is well trained, we train the image denoising network.
The input to the denoising network is an image \(I_{\text{noise}}\) corrupted by applying various noise masks \(M_{\text{noise}}\).
During the training phase, both \(I_{\text{noise}}\) before denoising and \(I_{\text{denoise}}\) after denoising are passed through the gradient-based XAI.
So, the corresponding medical image features \(F_{\text{denoised}}\) and \(F_{\text{clean}}\) are extracted.
Subsequently, the network minimizes the feature differences between \(F_{\text{denoised}}\) and \(F_{\text{clean}}\). 
This allows critical medical features to be preserved in the denoising process.
With \(F_{\text{denoised}}\) and \(F_{\text{clean}}\) are been visualized, we also allow medical experts to explain and trust the model outputs.
More implementation details will be covered in the following subsections.

\subsection{Gradient-based XAI and Sensitivity to Noise}
\label{gradient_based_xai}
Gradient-based XAI methods focus on utilizing gradient information from neural networks to explain model decisions for a given input.
Saliency Maps \cite{simonyan2013deep}, Grad-CAM \cite{selvaraju2017grad}, Guided Back-propagation \cite{springenberg2015striving}, etc. are typical methods for Gradient-based XAI.
While these methods are slightly different, their basic idea is to calculate the gradient of the model output relative to the input.
The calculated gradient is then used to quantify how much the input features affect the output.
Mathematically, for a neural network $f$ and a input $x$.
Gradient-based XAI computes \(\frac{\partial f(x)}{\partial x}\) to explain the decision of the model.
Although this method is simple and intuitive, it potentially faces some problems, such as sensitivity to noise.

To demonstrate the sensitivity, we consider a layer of neural network:
\begin{equation}
	f(x) = \sigma(w \cdot x + b),
\end{equation}
where \( \sigma \) is the Sigmoid activation function. \( w \) and \( b \) are the weight and bias. $x$ is the input.
We slightly add some noise to $x$.
So, the new input will be \( x' = x + \epsilon \).
In this case, the result of neural network will be $f(x') = \sigma(w \cdot (x + \epsilon) + b)$.
Now, we compute the gradient for $x$ and $x'$. 
Since derivative of the Sigmoid function is $\sigma'(x) = \frac{d}{dx}\left(\frac{1}{1 + e^{-x}}\right)=\frac{e^{-x}}{(1 + e^{-x})^2}=\sigma(x) \cdot (1 - \sigma(x))$, we can clearly see that even a slightly $\epsilon$ will lead the big change in $\sigma'(x)$ if $x$ closes to 0.
Also, this change might be amplified by $w$ since the gradient of the input is computed as $\frac{df}{dx'} = w \cdot \sigma'(w \cdot (x + \epsilon) + b)$.

Although this property is seen as a drawback in most cases, we are exploiting it to force the network to learn the difference between medical image features with and without noise.
Therefore, for the choice of gradient-based XAI methods, we did not choose methods such as SmoothGrad \cite{smilkov2017smoothgrad} or Integrated Gradients \cite{sundararajan2017axiomatic}, which suppress noise sensitivity.
We choose Guided Back-Propagation in this work.
In standard backpropagation, gradients are automatically propagated through the computational graph. 
However, in Guided Backpropagation, the propagation of the gradient is constrained, especially in the case of ReLU activation functions.
Mathematically, assuming that \( R \) is the gradient propagated from the previous layer, the Guided Backpropagation update in case of ReLU activation as follows:
\begin{equation}
	R_{\text{updated}} = R \odot \left( \frac{\partial f(x)}{\partial x} > 0 \right) \odot \left( x > 0 \right),
\end{equation}
where $R$ is the original gradients, $ \frac{\partial f(x)}{\partial x}$ is the gradient with respect to the input $x$, $\odot$ is the element-wise multiplication. 
$\frac{\partial f(x)}{\partial x} > 0$ is an indicator, determining whether the model is positively or negatively activated.
$x > 0$ determines the input $x$ is positive or negative.
Thus, during backpropagation, only gradients satisfy both positive model activations and inputs are retained.
In this way, the updated importance scores will contain only those features that contribute positively to the model predictions.
By visualizing this, an intuitive and useful explanation can be provided.

\subsection{Feature Preserving Loss}
Based on the discussion of Section \ref{gradient_based_xai}, we can assume a differentiable feature extraction method $\phi$ that utilizes the Guided Back-propagation.
For a corrupted medical image $I_{\text{noise}} \in \mathbb{R}^{h \times w}$, it can be considered as an overlay of a noise mask $M_{\text{noise}}$ and a clean image $I_{\text{clean}}$.
Mathematically, it can be written as $I_{\text{noise}} = M_{\text{noise}} + I_{\text{clean}}$.
Thus, for a pixel $I_{\text{noise}}(x,y)$ that locates at $I_{\text{noise}}$ at position $x$ and $y$, there are three possible cases.

\textbf{1)} The pixel is only consist by noise, namely $I_{\text{noise}}(x,y) = M_{\text{noise}}(x,y)$.

\textbf{2)} The pixel is only consist by medical feature, namely $I_{\text{noise}}(x,y) = I_{\text{clean}}(x,y)$.

\textbf{3)} The pixel is consist with noise and medical feature, namely $I_{\text{noise}}(x,y) = M_{\text{noise}}(x,y) + I_{\text{clean}}(x,y)$.

Let's recall the concept of feature importance weight $F \in \mathbb{R}^{h \times w}$ derived from gradient-based XAI, specifically from guided back-propagation. For each pixel at $(x, y)$, $F(x, y)$ quantifies the contribution of that pixel to a specific medical feature in the medical image.

The feature preserving loss, denoted as $L_{\text{FP}}$, can be defined mathematically as follows:

\begin{equation}
	L_{\text{FP}} = \sum_{x, y}  \lVert F_{\text{denoised}}(x, y) - F_{\text{clean}}(x, y) \rVert^{2},
\end{equation}
where $F_{\text{denoised}}$ is the output of guided backpropagation for denoised image, $F_{\text{clean}}$ is the output of guided backpropagation for clean image, and $\lVert \cdot \rVert^{2}$ denotes the squared Euclidean norm. 

Traditional loss functions and training methods such as mean square error (MSE) and residual learning work well for the first two cases (purely noisy and purely medical feature pixels).  
This is because the mean square error helps the neural network to globally model and learn the pattern and distribution of the noise.
However, when traditional denoising methods encounter the third case, both MSE and residual learning are not specifically designed to distinguish between features and noise within the same pixel.
They minimize the global difference between denoised and clean images, but this often comes at the cost of losing local features that are critical for medical diagnosis.
This is because of the convolutional operations that may be used by neural networks in optimizing MSE.
Especially in medical images, adjacent pixels tend to have similar brightness and color. 
In the process of minimizing MSE, the model will tend to maintain this local consistency and therefore may smooth the medical features.

The feature preserving loss makes the neural network more concerned about the feature weights on the medical image features. 
Since the feature preserving loss is obtained by using gradient-based explainable method, it takes into account the importance of each pixel independently. 
Thus making the neural network more focused on local features rather than global smoothing.

\subsection{Network Structure}
We use the two identical U-Net network architectures for gradient-based XAI feature extraction and image denoising. Since the U-Net architecture \cite{ronneberger2015u} is not our contribution, we discuss it briefly.
U-net consists of a symmetric encoder-decoder structure and an output convolutional layer. 
The input data is first passed through a feature extraction layer  using a $3\times3$ convolutional kernel and keeping the number of output channels at 64. Subsequently, the encoder starts downsampling through four `Down' modules, where each `Down' module consists of a maximum pooling layer and a double convolution, activated by ReLu. 
The number of channels is gradually increased from 64 to 128, 256, and 512 to capture higher-level semantic information.

The output of the encoder is then passed to the decoder which is upsampled by four `Up' modules. 
In this process, the feature maps corresponding to the layers of the encoder are concatenated with the output of the decoder. The number of channels is gradually reduced from 512 to 256, 128, 64, and finally 1 output channel is generated at the output layer by a $1\times1$ convolutional kernel.

Except feature preserving loss $L_{FP}$, we also applied the classic residual loss function for the purpose of denosing non-feature loss. Residual loss is commonly used to quantify the difference between the network output and the original clean signal, taking into account the noise component.
Specifically, we have noisy image $I_{\text{noise}}$, corresponding clean image $I_{\text{clean}}$. $I_{\text{denoised}}$ is the output of the denoising module. The residual \( r \) is defined as the difference between the denoised image and the noisy image:
\begin{equation}
	r = I_{\text{noise}} - I_{\text{denoised}}
\end{equation}

The residual loss $L_{res}$ is defined as the square of the difference between the residual \( r \) and the actual noise mask $M_{noise}$.
\begin{equation}
	L_{res}(r, M_{\text{noise}}) = \| r - M_{\text{noise}} \|^2,
\end{equation}

\section{Experimental Result}
In this section, we first provide the overview about the medical imaging dataset, applied noising types, evaluation metrics and implementation details. Then, we demonstrate the effectiveness of proposed feature preserving loss quantitatively and qualitatively.
Moreover, we conclude a in-depth discussion about the explainability that been provided by feature preserving loss.

\subsection{Medical Imaging Dataset}
To demonstrate the robustness of proposed loss function towards different acquisition equipments. Three public medical imaging dataset are been used.
They are Lung Image Database Consortium image collection (LIDC-IDRI) \cite{armato2011lung}, RSNA Pneumonia Detection \cite{wang2017chestx} and LiTS - Liver Tumor Segmentation \cite{bilic2023liver}.

\textbf{LIDC-IDRI:} Lung Image Database Consortium Image Collection (LIDC-IDRI) \cite{armato2011lung} is a publicly available medical image database. It focuses on early diagnosis and detection of lung cancer. LIDC-IDRI includes 1018 cases totaling more than 20,000 Computed Tomography (CT) images. 
Each case typically includes a series of slice images that are used to form a three-dimensional representation of the lung structure. 
These images are stored in DICOM (Digital Imaging and Communications in Medicine) format.

\textbf{RSNA:} RSNA Pneumonia Detection \cite{wang2017chestx} is a dataset used in medical image analysis competitions. The dataset typically consists of about 30,000 chest X-ray images that are categorized as either positive or negative for pneumonia.

\textbf{LiTS:} LiTS (Liver Tumor Segmentation) \cite{bilic2023liver} is a medical imaging database dedicated to liver and liver tumor segmentation.
LiTS datasets typically contain approximately 130 liver cases, each consisting of a series of consecutive computed tomography (CT) slices.

For the purpose of comparison, we randomly choose 5,000 images from each dataset for training. The images are reshaped to $(256 \times 256)$. Then, 500 images are randomly selected for testing purpose.

\subsection{Types of Synthesized Noise and Artifacts}
\begin{figure*}[tb]
	\includegraphics[width=\linewidth]{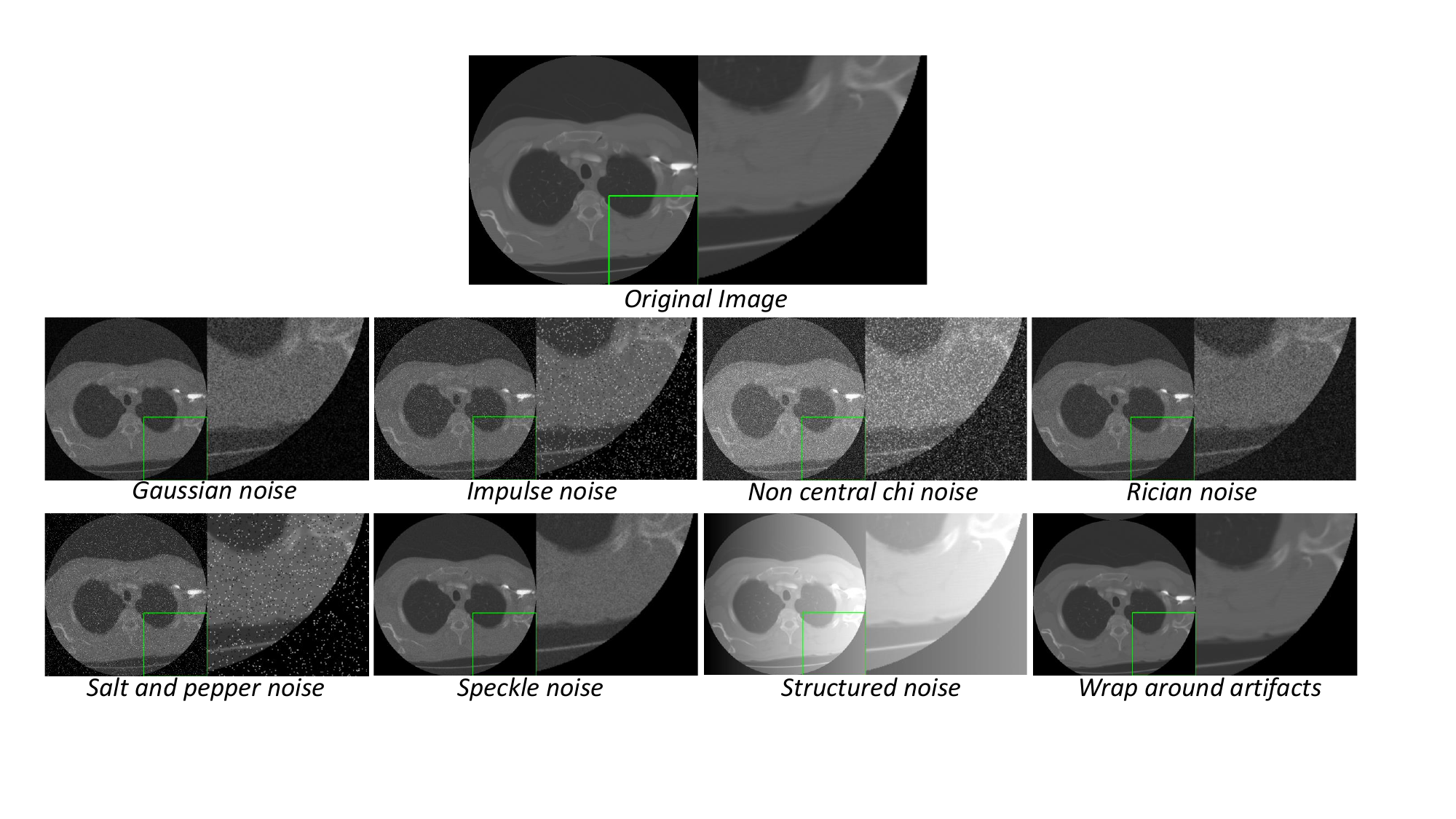}
	\vspace{-20pt}
	\caption{\label{noise_type}
		Sample images for synthesized noises.}
\end{figure*}

To evaluate the robustness of the proposed Feature Preserving Loss, we employ 13 common noise and synthetic artifacts in the medical image. Figure \ref{noise_type} shows the some of the sample noises. Also, Table \ref{tab:noise_artifacts} summarizes the 13 different noises and artifacts that applied in our experiment. 

\begin{table}[h]
	\centering
	\caption{Summary of Noise and Artifacts Applied in our Experiment.}
	\label{tab:noise_artifacts}
	\begin{threeparttable}
		\scalebox{0.9}{
			\begin{tabular}{|p{3cm}|p{5cm}|}
				\hline
				\textbf{Noise or Artifact}               & \textbf{Description} \\ \hline
				Chemical Shift (Chem. Shift)                & An image artifact caused by differences in the chemical composition of different tissues. \\ \hline
				Gaussian Noise (Gauss.)                 & A common type of image noise whose noise obeys a Gaussian distribution. \\ \hline
				Magnetic Field Inhomogeneity (Mag. Field)   & An image artifact due to uneven distribution of the magnetic field. \\ \hline
				Motion Artifacts   (Motion Art.)              & Image artifacts caused by patient or organ motion during the scanning process. \\ \hline
				Non-central Chi Noise  (Non-cent. Chi)         & A type of noise commonly found in magnetic resonance imaging (MRI). \\ \hline
				Poisson Noise                  & Usually arises due to the inherent variability of photon counting. \\ \hline
				Rician Noise (Rician)                   & A common noise in MRI images caused by fluctuations in the amplitude and phase of the signal. \\ \hline
				Salt and Pepper Noise (Salt \& Pepper)           & Usually appears as random white and black dots in the image. \\ \hline
				Speckle Noise                  & Noise due to interference of coherent light and is commonly seen in ultrasound images. \\ \hline
				Structured Noise (Struct. Noise)                & Usually generated by imperfections in scanning equipment or environmental disturbances. \\ \hline
				Susceptibility Artifacts (Susc. Art.)       & Result from inhomogeneities in the magnetic field or differences in the magnetization of a substance. \\ \hline
				Thermal Noise (Thermal)                  & Random noise due to the thermal movement of electrons. Thermal noise may reduce the signal-to-noise ratio of an image, affecting the clarity of the image. \\ \hline
				Wrap-around Artifacts (Wrap Around)          & Caused by phase distortion during image reconstruction. \\ \hline
		\end{tabular}}
		\begin{tablenotes}
			\small
			\item The contents of the parentheses will be used as abbreviations for Tables \ref{tab:comparison_LIDC}, \ref{tab:comparison_LiTs} and \ref{tab:comparison_RSNA}.
		\end{tablenotes}
	\end{threeparttable}
\end{table}

\subsection{Evaluation Metric}
Similar to other regular denoising metrics, we use Peak Signal-to-Noise Ratio (PSNR) and Structural Similarity Index (SSIM) as evaluation metrics. 
The inputs for the computing PSNR and SSIM are the original image $I_{\text{clean}}$ and the denoised image $I_{\text{denoised}}$.

PSNR is a metric used to measure the difference between the original image and the denoised image. The higher its value, the better the denoising effect. PSNR is defined as 
\begin{equation}
	\text{PSNR} = 10 \cdot \log_{10}\left(\frac{{{\text{MAX}_I^2}}}{{\text{MSE}}}\right),
\end{equation}
where \(\text{MAX}_I\) is the maximum possible pixel value of the image (e.g., for an 8-bit image, \(\text{MAX}_I = 255\)), and \(\text{MSE}\) is the mean-square error, computed as
\begin{equation}
	\text{MSE} = \frac{1}{{m \cdot n}} \sum_{i=0}^{m-1} \sum_{j=0}^{n-1} \left( I_{\text{clean}}(i,j) - I_{\text{denoised}}(i,j) \right)^2,
\end{equation}
where \(m\) and \(n\) are the dimensions of the image.

SSIM is used to measure the structural similarity of two images. SSIM takes into account the brightness, contrast and structural information. The value of SSIM ranges from -1 to 1. The closer the value is to 1, the better the denoising effect. SSIM is defined as 
\begin{equation}
	\text{SSIM}(x, y) = \frac{(2\mu_x \mu_y + c_1)(2\sigma_{xy} + c_2)}{(\mu_x^2 + \mu_y^2 + c_1)(\sigma_x^2 + \sigma_y^2 + c_2)},
\end{equation}
where \(x\) and \(y\) are the $I_{\text{clean}}$ and $I_{\text{denoised}}$. \(\mu_x\) and \(\mu_y\) are the mean values of the images. \(\sigma_x^2\) and \(\sigma_y^2\) are the variances of the images, \(\sigma_{xy}\) is the covariance of the image, \(c_1\) and \(c_2\) are constants to prevent the denominator from being zero.

\subsection{Quantitative Evaluation}
\label{quan}
Tables \ref{tab:comparison_LIDC}, \ref{tab:comparison_LiTs} and \ref{tab:comparison_RSNA} demonstrate the evaluate results for dataset LIDC, LiTs and RSNA.
We compare our method with 9 common denoising methods under different noise and artifact conditions.
To save space on the table, we use abbreviations.
The correspondence is DPIR: Deep Plug-and-Play Image Restoration; Bilat.: Bilateral Filter; Filt. Comb.: Combined denoising strategies by Mean Filter, Median Filter and Gaussian Filter; Gaus. Filt.: Gaussian Filter; Mean Filt.: Mean Filter; Medi Filt.: Median Filter; NL Mean.: Non-local Means Denoising. Wave. Den.: Wavelet Denoising; Wien. Filt.: Wiener Filter.

\begin{table*}[htbp]
	\centering
	\caption{LIDC PSNR and SSIM Comparison}
	\label{tab:comparison_LIDC}
	\begin{threeparttable}
		\scalebox{1}{
			\begin{tabular}{|c|p{2cm}|p{1cm}|p{1cm}|p{1cm}|p{1cm}|p{1cm}|p{1cm}|p{1cm}|p{1cm}|p{1cm}|p{1cm}|p{1cm}|}
				\hline 
				& \textbf{Noise Type} & \textbf{Our} & \textbf{DPIR} & \textbf{Bilat.} & \textbf{Filt. Comb.} & \textbf{Gaus. Filt.} & \textbf{Mean Filt.} & \textbf{Medi. Filt.} & \textbf{NL Mean.} & \textbf{Wave. Den.} & \textbf{Wien. Filt.} \\ \hline
				\multirow{15}{*}{\textbf{PSNR}}&Chem. Shift & \textbf{49.39} & 35.95 & 36.01 & 35.70 & 35.82 & 35.74 & 35.99 & 35.65 & 35.61 & 34.79 \\ \cline{2-12}
				&Gauss. & \textbf{35.20} & 30.74 & 30.92 & 32.14 & 30.92 & 31.10 & 31.37 & 31.69 & 31.75 & 31.46 \\ \cline{2-12}
				&Mag. Field & \textbf{55.30} & 31.49 & 31.45 & 31.38 & 31.69 & 31.54 & 31.71 & 31.40 & 31.94 & 30.81 \\ \cline{2-12}
				&Motion Art. & \textbf{37.20} & 32.80 & 32.69 & 32.49 & 32.47 & 32.47 & 32.55 & 32.90 & 32.34 & 32.11 \\ \cline{2-12}
				&Non-cent. Chi & \textbf{32.17} & 27.55 & 27.53 & 27.28 & 27.38 & 27.36 & 27.47 & 27.92 & 27.34 & 28.19 \\ \cline{2-12}
				&Poisson Noise & 35.25 & \textbf{36.97} & 34.52 & 35.53 & 34.24 & 34.45 & 33.67 & 36.34 & 34.16 & 32.46 \\ \cline{2-12}
				&Rician & \textbf{33.88} & 29.73 & 29.32 & 30.35 & 29.62 & 29.75 & 29.45 & 30.50 & 30.30 & 30.61 \\ \cline{2-12}
				&Salt \& Pepper & 38.60 & 36.08 & 36.68 & 32.60 & 33.18 & 32.53 & 41.10 & 38.86 & \textbf{41.50} & 31.73 \\ \cline{2-12}
				&Speckle Noise & 36.36 & \textbf{37.78} & 35.94 & 36.15 & 35.73 & 35.68 & 35.17 & 36.96 & 34.66 & 33.36 \\ \cline{2-12}
				&Struct. Noise & \textbf{30.54} & 27.99 & 27.98 & 27.97 & 27.97 & 27.97 & 27.97 & 28.01 & 27.99 & 28.18 \\ \cline{2-12}
				&Susc. Art. & \textbf{45.21} & 30.09 & 30.12 & 30.08 & 30.15 & 30.11 & 30.17 & 30.09 & 30.23 & 29.77 \\ \cline{2-12}
				&Thermal & \textbf{34.96} & 30.74 & 30.92 & 32.14 & 30.92 & 31.10 & 31.37 & 31.69 & 31.75 & 31.46 \\ \cline{2-12}
				&Wrap Around & \textbf{35.26} & 31.57 & 31.48 & 31.35 & 31.35 & 31.34 & 31.41 & 31.60 & 31.28 & 31.12 \\ 
				\hline
				\multirow{15}{*}{\textbf{SSIM}}&Chem. Shift & \textbf{0.98} & 0.83 & 0.83 & 0.84 & 0.83 & 0.83 & 0.82 & 0.82 & 0.81 & 0.66 \\ \cline{2-12}
				&Gauss. & \textbf{0.82} & 0.50 & 0.50 & 0.65 & 0.53 & 0.55 & 0.55 & 0.51 & 0.60 & 0.35 \\ \cline{2-12}
				&Mag. Field & \textbf{0.99} & 0.87 & 0.88 & 0.88 & 0.92 & 0.89 & 0.90 & 0.86 & 0.96 & 0.68 \\ \cline{2-12}
				&Motion Art. & \textbf{0.86} & 0.64 & 0.65 & 0.64 & 0.63 & 0.63 & 0.63 & 0.65 & 0.60 & 0.52 \\ \cline{2-12}
				&Non-cent. Chi & \textbf{0.76} & 0.13 & 0.14 & 0.45 & 0.30 & 0.32 & 0.29 & 0.25 & 0.41 & 0.26 \\ \cline{2-12}
				&Poisson Noise & \textbf{0.91} & 0.87 & 0.81 & 0.85 & 0.80 & 0.80 & 0.76 & 0.87 & 0.79 & 0.54 \\ \cline{2-12}
				&Rician & \textbf{0.81} & 0.46 & 0.45 & 0.60 & 0.49 & 0.50 & 0.45 & 0.50 & 0.56 & 0.37 \\ \cline{2-12}
				&Salt \& Pepper & \textbf{0.88} & 0.29 & 0.27 & 0.63 & 0.46 & 0.47 & 0.92 & 0.32 & 0.29 & 0.31 \\ \cline{2-12}
				&Speckle Noise & \textbf{0.91} & 0.89 & 0.86 & 0.87 & 0.84 & 0.84 & 0.81 & 0.88 & 0.82 & 0.59 \\ \cline{2-12}
				&Struct. Noise & \textbf{0.83} & 0.55 & 0.55 & 0.56 & 0.58 & 0.56 & 0.57 & 0.53 & 0.61 & 0.50 \\ \cline{2-12}
				&Susc. Art. & \textbf{0.96} & 0.80 & 0.80 & 0.81 & 0.84 & 0.82 & 0.83 & 0.79 & 0.87 & 0.58 \\ \cline{2-12}
				&Thermal & \textbf{0.83} & 0.50 & 0.50 & 0.65 & 0.53 & 0.55 & 0.55 & 0.51 & 0.60 & 0.35 \\ \cline{2-12}
				&Wrap Around & \textbf{0.82} & 0.59 & 0.60 & 0.59 & 0.58 & 0.58 & 0.58 & 0.60 & 0.56 & 0.48 \\ \hline
				
			\end{tabular}
		}
		\begin{tablenotes}
			\small
			\item The full names of the abbreviations can be found in Table \ref{tab:noise_artifacts} and Section \ref{quan}.
		\end{tablenotes}
	\end{threeparttable}
\end{table*}

Among them, DPIR \cite{zhang2021plug} is the deep learning advanced-performed denoising algorithm. 
In three different datasets, DPIR is the best method other than our proposed method in most cases.
Although the DPIR algorithm has considered Noise Level Map for preprocessing, this approach still has limitations in processing medical images.
Specifically, it is difficult for the Noise Level Map to capture various local features and the corresponding feature intensities in medical images.
This causes the algorithm perform poorly when dealing with noises that are particularly destructive to medical features, such as non-central Chi noise.
Compare with DPIR, our proposed Feature Preserving Loss is better.
Feature Preserving Loss not only adaptively learns key features related to medical image reconstruction during the network training, but also quantifies the changes in features before and after denoising.
Subsequently, by minimizing these changes, we can ensure that more important feature information for medical diagnostic can be retained during the denoising process.
\begin{table*}[tbp]
	\centering
	\caption{LiTs PSNR and SSIM Comparison}
	\label{tab:comparison_LiTs}
	\begin{threeparttable}
		\scalebox{1}{
			\begin{tabular}{|c|p{2cm}|p{1cm}|p{1cm}|p{1cm}|p{1cm}|p{1cm}|p{1cm}|p{1cm}|p{1cm}|p{1cm}|p{1cm}|p{1cm}|}
				\hline 
				& \textbf{Noise Type} & \textbf{Our} & \textbf{DPIR} & \textbf{Bilat.} & \textbf{Filt. Comb.} & \textbf{Gaus. Filt.} & \textbf{Mean Filt.} & \textbf{Medi. Filt.} & \textbf{NL Mean.} & \textbf{Wave. Den.} & \textbf{Wien. Filt.} \\ \hline
				\multirow{15}{*}{\textbf{PSNR}}& Chem. Shift& \textbf{51.67}& 36.57& 36.83& 36.47& 36.37& 36.40& 36.56& 36.33& 36.15& 34.36 \\ \cline{2-12}
				& Gauss.& \textbf{37.40}& 31.13& 31.37& 31.99& 31.00& 31.15& 32.01& 31.70& 31.66& 32.24 \\ \cline{2-12}
				& Mag. Field& \textbf{55.12}& 34.02& 33.79& 33.65& 34.24& 34.00& 34.50& 33.77& 34.50& 32.15 \\ \cline{2-12}
				& Motion Art.& \textbf{39.58}& 34.36& 34.25& 34.01& 34.01& 34.01& 34.12& 34.37& 33.97& 32.82 \\ \cline{2-12}
				& Non-cent. Chi& \textbf{34.41}& 27.57& 27.55& 27.27& 27.38& 27.36& 27.45& 28.03& 27.35& 28.44 \\ \cline{2-12}
				& Poisson Noise& 39.41& \textbf{39.70}& 36.18& 37.16& 36.08& 36.30& 35.47& 37.77& 34.18& 33.42 \\ \cline{2-12}
				& Rician& \textbf{35.09}& 29.19& 28.66& 29.33& 28.91& 29.00& 28.82& 29.80& 29.18& 30.48 \\ \cline{2-12}
				& Salt \& Pepper& \textbf{44.94}& 37.51& 38.46& 32.93& 34.19& 33.33& 44.07& 40.70& 42.83& 31.87 \\ \cline{2-12}
				& Speckle Noise& 39.84& \textbf{40.21}& 37.76& 37.80& 37.79& 37.67& 37.35& 38.93& 35.05& 34.12 \\ \cline{2-12}
				& Struct. Noise& \textbf{31.23}& 28.09& 28.08& 28.07& 28.07& 28.07& 28.07& 28.13& 28.10& 28.26 \\ \cline{2-12}
				& Susc. Art.& \textbf{47.52}& 33.03& 33.01& 32.95& 33.29& 33.15& 33.32& 32.88& 33.37& 31.90 \\ \cline{2-12}
				& Thermal& \textbf{36.99}& 31.13& 31.37& 31.99& 31.00& 31.15& 32.01& 31.70& 31.66& 32.24 \\ \cline{2-12}
				& Wrap Around& \textbf{36.39}& 33.10& 33.01& 32.84& 32.86& 32.85& 32.95& 33.09& 32.86& 32.00 \\
				\hline
				\multirow{15}{*}{\textbf{SSIM}} & Chem. Shift& \textbf{0.94}& 0.86& 0.87& 0.88& 0.86& 0.87& 0.86& 0.85& 0.84& 0.48 \\ \cline{2-12}
				& Gauss.& \textbf{0.85}& 0.52& 0.51& 0.64& 0.52& 0.55& 0.59& 0.45& 0.57& 0.28 \\ \cline{2-12}
				& Impulse& \textbf{0.89}& 0.30& 0.29& 0.62& 0.44& 0.46& 0.98& 0.32& 0.30& 0.25 \\ \cline{2-12}
				& Mag. Field& \textbf{0.99}& 0.92& 0.92& 0.93& 0.96& 0.95& 0.95& 0.89& 0.95& 0.52 \\ \cline{2-12}
				& Motion Art.& \textbf{0.85}& 0.76& 0.76& 0.75& 0.74& 0.74& 0.74& 0.75& 0.72& 0.40 \\ \cline{2-12}
				& Non-cent. Chi& \textbf{0.76}& 0.10& 0.11& 0.37& 0.24& 0.26& 0.24& 0.16& 0.32& 0.20 \\ \cline{2-12}
				& Poisson Noise& \textbf{0.94}& 0.93& 0.87& 0.92& 0.87& 0.88& 0.84& 0.92& 0.79& 0.44 \\ \cline{2-12}
				& Rician& \textbf{0.79}& 0.41& 0.39& 0.52& 0.42& 0.44& 0.40& 0.41& 0.47& 0.31 \\ \cline{2-12}
				& Salt \& Pepper& \textbf{0.93}& 0.30& 0.29& 0.62& 0.44& 0.46& 0.98& 0.32& 0.30& 0.25 \\ \cline{2-12}
				& Speckle Noise& \textbf{0.95}& 0.94& 0.92& 0.93& 0.91& 0.91& 0.89& 0.92& 0.82& 0.47 \\ \cline{2-12}
				& Struct. Noise& \textbf{0.85}& 0.48& 0.48& 0.49& 0.51& 0.50& 0.50& 0.47& 0.52& 0.41 \\ \cline{2-12}
				& Susc. Art.& \textbf{0.98}& 0.85& 0.86& 0.86& 0.88& 0.88& 0.88& 0.82& 0.87& 0.47 \\ \cline{2-12}
				& Thermal& \textbf{0.84}& 0.52& 0.51& 0.64& 0.52& 0.55& 0.59& 0.45& 0.57& 0.28 \\ \cline{2-12}
				& Wrap Around& \textbf{0.85}& 0.70& 0.70& 0.70& 0.69& 0.69& 0.69& 0.70& 0.67& 0.37 \\  \hline
				
			\end{tabular}
		}
		\begin{tablenotes}
			\small
			\item The full names of the abbreviations can be found in Table \ref{tab:noise_artifacts} and Section \ref{quan}.
		\end{tablenotes}
	\end{threeparttable}
\end{table*}

In contrast, the classical denoising algorithms, such as Wiener filters, median filters or Gaussian filters, have difficulty in maintaining denoising stability under different noise or artifacts conditions.
This is due to the fact that classical denoising algorithms mainly rely on mathematical models to describe the relationship between images and noise.
Therefore, they are usually manually designed for specific types of noise.
In the case of the Wiener filter, the Wiener filter assumes that the image and the noise are linearly additive and tries to minimize the prediction error to estimate the original image.
Thus, the Wiener filter tends to perform well when targeting a specific type of noise, such as salt and pepper noise.
However, the Wiener filter becomes less effective at denoising when dealing with more complex or diverse noises.

Compared with other famous denosing algorithms, the superiority of our proposed method is demonstrated in the Structural Similarity Index (SSIM).
SSIM assesses image quality from the perspective of human visual perception. 
On three different datasets, our method performs well on SSIM, ahead of other algorithms. 
This result not only validates the effectiveness of our loss function in feature preserving, but also shows that the method has a clear advantage in structural quality.

\begin{table*}[tbp]
	\centering
	\caption{RSNA PSNR and SSIM Comparison}
	\label{tab:comparison_RSNA}
	\begin{threeparttable}
		\scalebox{1}{
			\begin{tabular}{|c|p{2cm}|p{1cm}|p{1cm}|p{1cm}|p{1cm}|p{1cm}|p{1cm}|p{1cm}|p{1cm}|p{1cm}|p{1cm}|p{1cm}|}
				\hline 
				& \textbf{Noise Type} & \textbf{Our} & \textbf{DPIR} & \textbf{Bilat.} & \textbf{Filt. Comb.} & \textbf{Gaus. Filt.} & \textbf{Mean Filt.} & \textbf{Medi. Filt.} & \textbf{NL Mean.} & \textbf{Wave. Den.} & \textbf{Wien. Filt.} \\ \hline
				\multirow{15}{*}{\textbf{PSNR}}& Chem. Shift& \textbf{41.99}& 35.07& 35.29& 35.18& 35.06& 35.09& 35.23& 34.57& 34.88& 33.61 \\ \cline{2-12}
				& Gauss.& \textbf{34.65}& 30.12& 30.11& 31.77& 30.38& 30.63& 30.26& 30.47& 31.23& 29.50 \\ \cline{2-12}
				& Mag. Field& \textbf{49.86}& 30.11& 30.12& 30.15& 30.25& 30.21& 30.28& 29.95& 30.31& 29.69 \\ \cline{2-12}
				& Motion Art.& \textbf{34.47}& 31.29& 31.22& 31.14& 31.11& 31.11& 31.12& 31.34& 31.03& 30.81 \\ \cline{2-12}
				& Non-cent. Chi& \textbf{34.05}& 27.85& 27.81& 27.57& 27.63& 27.61& 27.79& 28.03& 27.66& 27.93 \\ \cline{2-12}
				& Poisson Noise& \textbf{35.80}& 32.47& 31.55& 33.03& 31.62& 31.89& 31.25& 32.91& 31.55& 29.82 \\ \cline{2-12}
				& Rician& \textbf{34.25}& 29.82& 29.73& 31.10& 30.03& 30.25& 29.86& 30.53& 30.77& 29.54 \\ \cline{2-12}
				& Salt \& Pepper& \textbf{41.62}& 33.36& 33.78& 31.43& 31.32& 31.07& 37.07& 34.15& 35.42& 30.19 \\ \cline{2-12}
				& Speckle Noise& \textbf{36.11}& 32.83& 32.09& 33.21& 32.13& 32.33& 31.82& 32.84& 31.20& 30.05 \\ \cline{2-12}
				& Struct. Noise& \textbf{30.34}& 28.00& 27.99& 27.98& 27.99& 27.98& 27.99& 28.00& 28.01& 27.99 \\ \cline{2-12}
				& Susc. Art.& \textbf{42.88}& 29.80& 29.81& 29.86& 29.97& 29.93& 30.01& 29.67& 30.10& 29.61 \\ \cline{2-12}
				& Thermal& \textbf{34.75}& 30.12& 30.11& 31.77& 30.38& 30.63& 30.26& 30.47& 31.23& 29.50 \\ \cline{2-12}
				& Wrap Around& \textbf{30.65}& 29.89& 29.85& 29.82& 29.81& 29.80& 29.81& 29.87& 29.74& 29.62 \\ 
				\hline
				\multirow{15}{*}{\textbf{SSIM}} & Chem. Shift& \textbf{0.98}& 0.89& 0.90& 0.90& 0.89& 0.89& 0.89& 0.88& 0.87& 0.77 \\ \cline{2-12}
				& Gauss.& \textbf{0.91}& 0.55& 0.55& 0.81& 0.63& 0.67& 0.60& 0.69& 0.78& 0.42 \\ \cline{2-12}
				& Mag. Field& \textbf{0.99}& 0.91& 0.91& 0.93& 0.96& 0.95& 0.95& 0.89& 0.97& 0.77 \\ \cline{2-12}
				& Motion Art.& \textbf{0.95}& 0.75& 0.75& 0.74& 0.72& 0.73& 0.72& 0.76& 0.71& 0.65 \\ \cline{2-12}
				& Non-cent. Chi& \textbf{0.91}& 0.22& 0.23& 0.66& 0.43& 0.48& 0.39& 0.44& 0.56& 0.36 \\ \cline{2-12}
				& Poisson Noise& \textbf{0.94}& 0.78& 0.72& 0.87& 0.74& 0.77& 0.71& 0.90& 0.77& 0.49 \\ \cline{2-12}
				& Rician& \textbf{0.90}& 0.55& 0.54& 0.79& 0.62& 0.66& 0.58& 0.70& 0.77& 0.43 \\ \cline{2-12}
				& Salt \& Pepper& \textbf{0.98}& 0.29& 0.26& 0.73& 0.50& 0.54& 0.97& 0.30& 0.25& 0.34 \\ \cline{2-12}
				& Speckle Noise& \textbf{0.95}& 0.81& 0.77& 0.89& 0.79& 0.81& 0.75& 0.90& 0.71& 0.54 \\ \cline{2-12}
				& Struct. Noise& \textbf{0.92}& 0.76& 0.76& 0.78& 0.79& 0.79& 0.79& 0.73& 0.80& 0.72 \\ \cline{2-12}
				& Susc. Art.& \textbf{0.99}& 0.82& 0.82& 0.84& 0.86& 0.85& 0.86& 0.80& 0.87& 0.72 \\ \cline{2-12}
				& Thermal& \textbf{0.92}& 0.55& 0.55& 0.81& 0.63& 0.67& 0.60& 0.69& 0.78& 0.42 \\ \cline{2-12}
				& Wrap Around& \textbf{0.91}& 0.69& 0.69& 0.68& 0.66& 0.67& 0.66& 0.70& 0.65& 0.60 \\   \hline
				
			\end{tabular}
		}
		\begin{tablenotes}
			\small
			\item The full names of the abbreviations can be found in Table \ref{tab:noise_artifacts} and Section \ref{quan}.
		\end{tablenotes}
	\end{threeparttable}
\end{table*}

\begin{figure}[htb]
	\includegraphics[width=\linewidth]{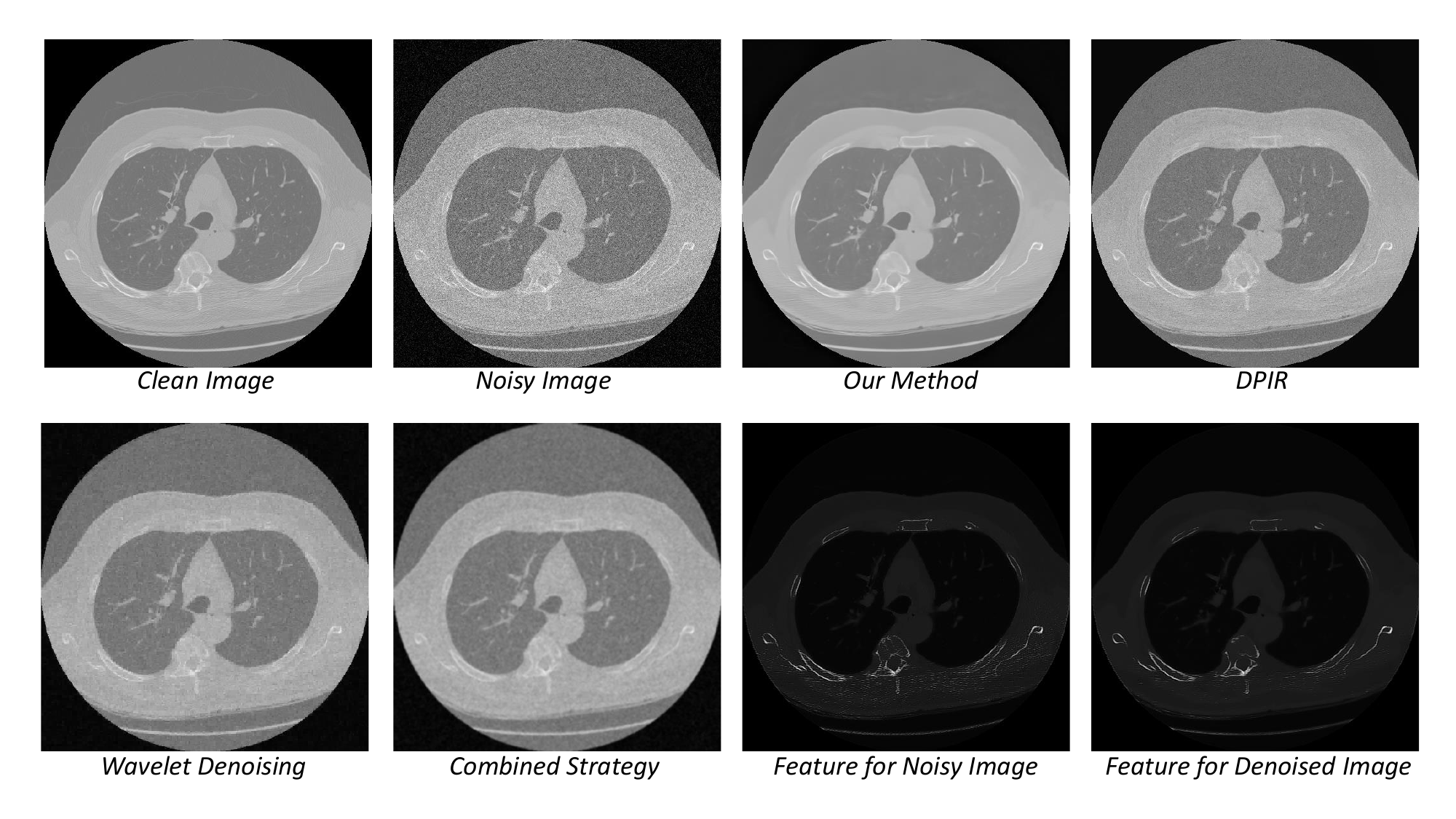}
	\caption{\label{vis_lidc}
		Denosing Visualization Result for LIDC dataset as Gaussian Noise Applied.}
\end{figure}

\begin{figure}[htb]
	\includegraphics[width=\linewidth]{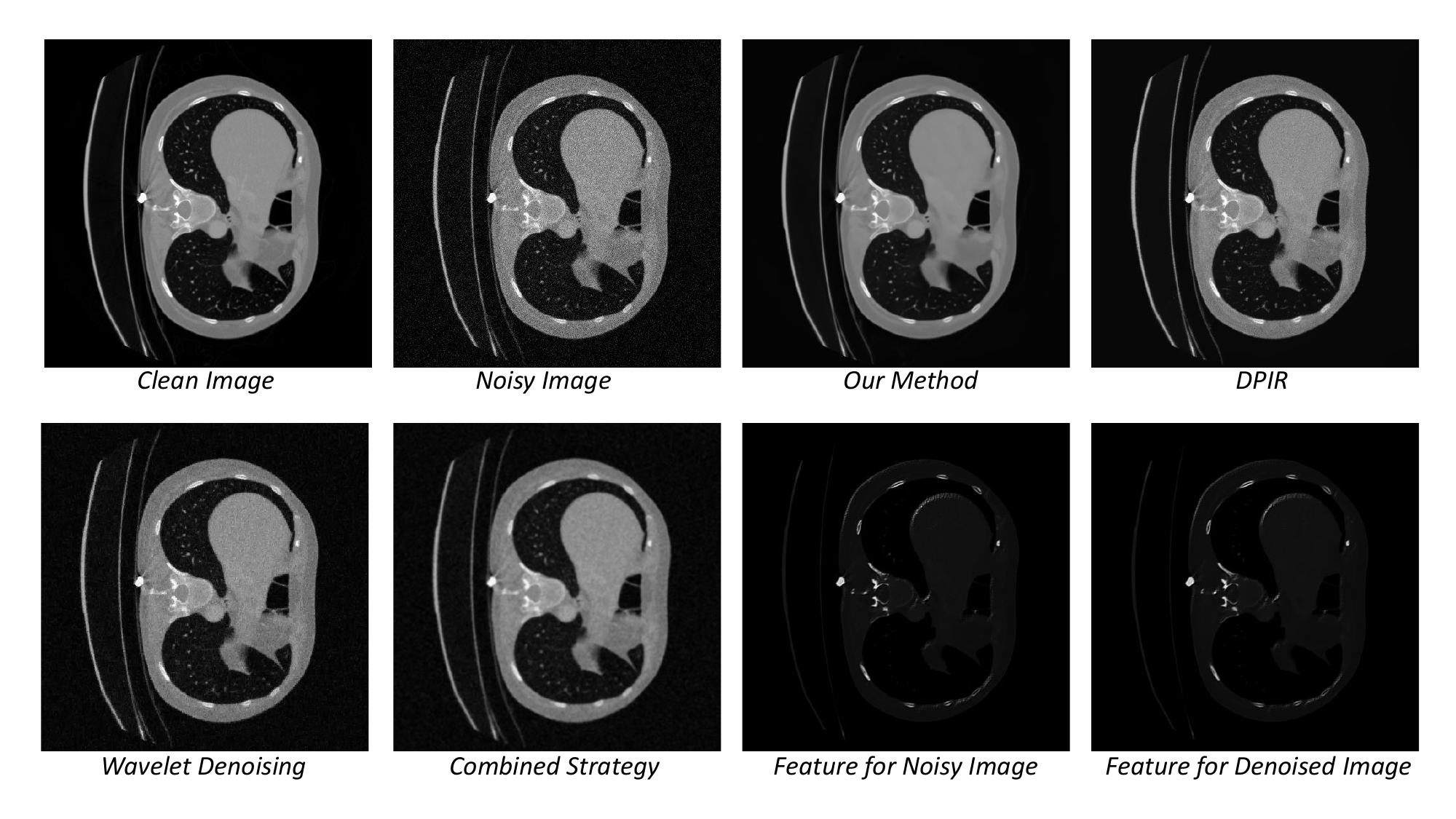}
	\caption{\label{vis_lits}
		Denosing Visualization Result for LiTs dataset as Gaussian Noise Applied.}
\end{figure}

\begin{figure}[htb]
	\includegraphics[width=\linewidth]{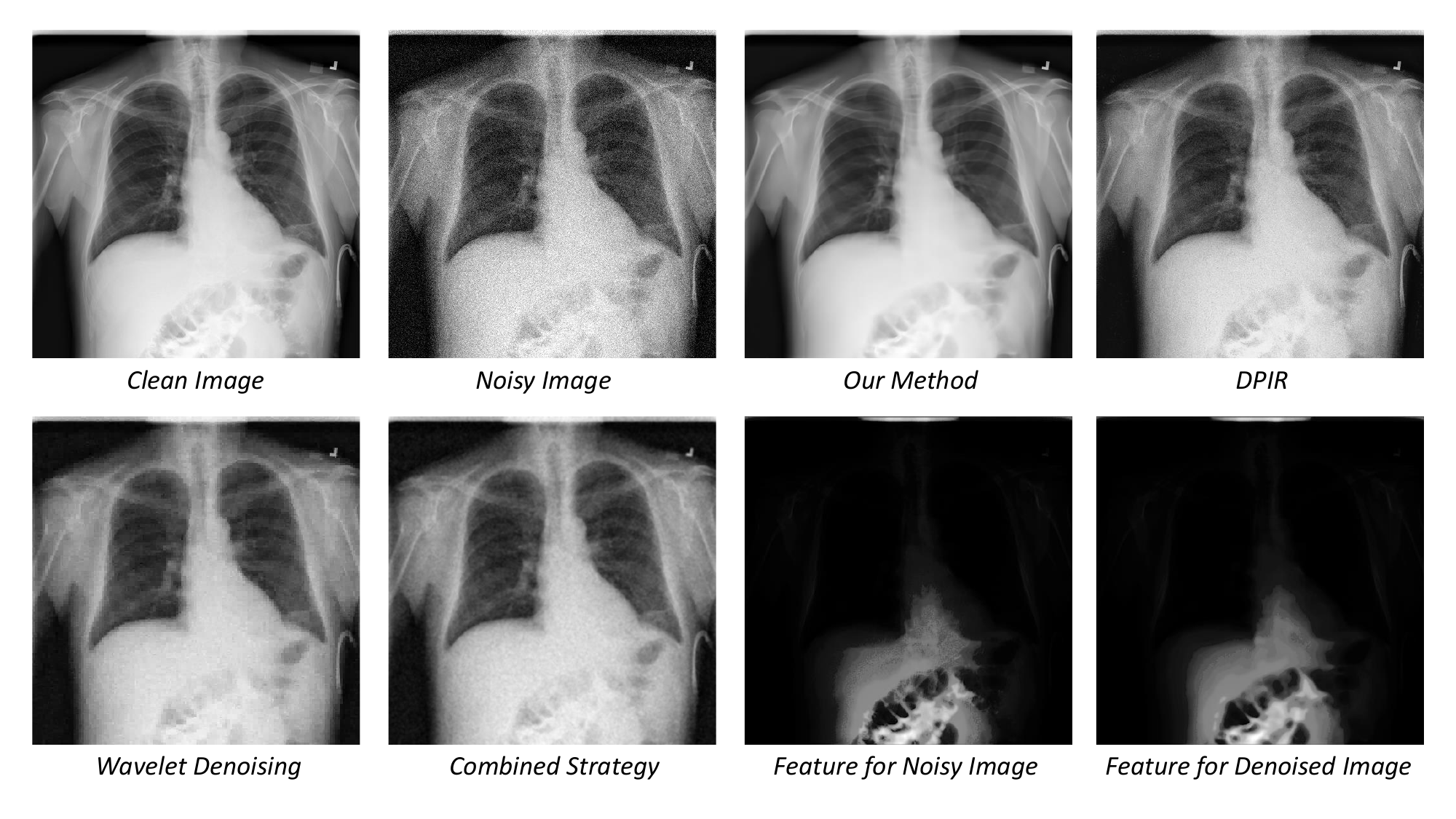}
	\caption{\label{vis_rsna}
		Denosing Visualization Result for RSNA dataset as Gaussian Noise Applied.}
\end{figure}

\subsection{Qualitative Evaluation}
Figures \ref{vis_lidc}, \ref{vis_lits} and \ref{vis_rsna} show the visual denoising results of our work. 
For each image, the first row from left to right shows: the clean image, the noisy image, our denoising result, and the DPIR denoising result. The second row from left to right are: result of Wavelet Denoising, result of Combine Strategy, feature of the noisy image, feature of the denoised image.
Due to the length limitation, the noise we show in the Figures \ref{vis_lidc}, \ref{vis_lits} and \ref{vis_rsna} is Gaussian noise. 
The complete denoising results can be found in Google Drive\footnote{https://drive.google.com/drive/folders/\newline17QxZauxFhdULcI1NpTDNNe8P4kx26Yab?usp=sharing}. 
We will also publicly release the code after the paper is accepted.

Overall, compared with other denoising algorithms, our method performs much better in terms of denoising performance. 
In particular, when compared with the DPIR, our method performs particularly well in terms of noise control.
Our denoising results show almost no presence of granular noise points.
Typically, the cost of eliminating granularity is that critical features in medical images may be over-smoothed or completely blurred. 
However, in our method, important feature regions in medical images can be effectively preserved with the help of Feature Preserving Loss.

As Figure \ref{vis_rsna} shows, for lung region, the DPIR algorithm result obviously lack a sense of hierarchy.
The entire lung region is almost shows as a uniform white color. 
In contrast, our method successfully preserves the layering of the lung region guided by the feature preservation loss. 
Although there is a small difference compared to the clean image, our results can be still considered very satisfactory.
Other visualization results also demonstrate the superiority of our method. 
As can be seen in Figure \ref{vis_lidc}, our method performs much better in preserving the features of blood vessels and lesion regions compared to other algorithms. 
Considering the severe damage to the image by noise, our method has great potential to solve the feature preservation problem in denoising.

\subsection{Discussion on Explainability}
It is worth to mention that we do not claim to propose a new explainable method for medical images. 
The ``medical image features'' mentioned in this paper may differ from those in a real medical environment. 
The features we used are extracted from the reconstructed network by gradient-based XAI method. 
These features quantify the importance of each input pixel in the reconstruction process. 
Although in most cases, the features extracted by XAI are consistent with those of the actual medical images (e.g., Figures \ref{vis_lidc}, \ref{vis_lits} and \ref{vis_rsna}). However, due to the lack of guiding information, there is also a possibility of bias between the two.

In fact, a more ideal approach may be to train a medical image feature detection network with human annotation. 
The importance of these features is then further quantified by a gradient XAI method. 
However, due to the lack of labeled datasets, using an existing reconstructed network is a practical compromise. 
Despite the possible differences in features between the two, our denoising results still demonstrate excellent performance and effective feature preserving ability. 
This further confirms that utilizing feature preserving loss is a correct strategy. 
In the future, we will focus on obtaining more accurate features.

\section{Conclusion}
In this work, we purposed a new loss function called Feature Preserving Loss that helps denosing network preserving medical features essential for diagnosis.
This new loss function exploits the capabilities of gradient-based eXplainable Artificial Intelligence (XAI) and leverages the sensitivity of gradient-based XAI methods to noise.
Notably, this approach not only excels in denoising but also enhances the explainability of the model by visualizing the feature map.
Future research can focus on obtaining more accurate medical features.
Overall, we present an effective and explainable denosing method, which has the great potential for application in medical image denoising.

\bibliographystyle{IEEEtran}  
\bibliography{ref}

\end{document}